\documentclass[a4paper,10pt,twoside]{cpc-hepnp}
\usepackage{multicol}
\usepackage{graphicx}
\usepackage{booktabs}
\usepackage{amssymb,bm,mathrsfs,bbm,amscd}
\usepackage{lastpage}
\usepackage{changepage} 

\usepackage[scientific-notation=true]{siunitx}
\usepackage{multirow}
\usepackage{tikz}
\usetikzlibrary{shapes.geometric, arrows}

\usepackage[utf8]{inputenc}
\usepackage[colorlinks=true, linkcolor=red, citecolor=blue]{hyperref}

\usepackage{amsmath,amsfonts,amssymb,bm}
\usepackage{graphicx}
\usepackage{color}
\usepackage{subfigure}
\usepackage{multirow}
\usepackage{textcomp}
\usepackage{slashed}
\usepackage{cancel}

\def\({\left(}
\def\){\right)}
\def\[{\left[}
\def\]{\right]}

\def\be{\begin{eqnarray}}
\def\ee{\end{eqnarray}}
\usepackage[capitalise]{cleveref}
\crefname{figure}{Fig.}{Figs.}
\Crefname{figure}{Fig.}{Figs.}

\begin{document}

\fancyhead[c]{\small Submitted to Chinese Physics C} \fancyfoot[C]{\small Page-\thepage}



\title{The Radial Excited Heavy Mesons
\footnote{This work is supported by: the National Natural Science Foundation of China under contracts No. 11947108 and No. 12005060.}}

\author{%
      Muyang Chen $^{1}$\email{muyang@hunnu.edu.cn}
}

\maketitle

\address{%
$^1$ Department of Physics, Hunan Normal University, Changsha 410081, China
}

\begin{abstract}
In this paper, the first radial excited heavy pseudoscalar and vector mesons ($\eta_c(2S)$, $\psi(2S)$, $B_c(2S)$, $B^*_c(2S)$, $\eta_b(2S)$, $\varUpsilon(2S)$) are studied in the Dyson-Schwinger equation and Bethe-Salpeter equation approach. It is showed that the effective interactions of the radial excited states are harder than that of the ground states. With the interaction well determined by fitting the masses and leptonic decay constants of $\psi(2S)$ and $\varUpsilon(2S)$, the first radial excited heavy mesons could be quantitatively described in the rainbow ladder approximation. The masses and leptonic decay constants of $\eta_c(2S)$, $B_c(2S)$, $B^*_c(2S)$ and $\eta_b(2S)$ are predicted.
\end{abstract}




\begin{multicols}{2}

\section{Introduction}
\label{sec:intro} 
\noindent

 For theoretical scientist, the radial excited hadrons are more challenging than the ground states. They are widely studied in the quark model \cite{Godfrey1985,Capstick1985,Ebert2009,Ebert2011,Song2015,Li2019}. However, many ``missing states" are predicted. A more rigorous corresponding relation between the quark model resonant states and the experimental observed ones is being expected. Lattice Quantum Chromodynamics (QCD) aims to produce the hadron spectrum from first principle. The study of radial excited states is also less advanced than the ground states. See table I and II in Ref. \cite{Chen2019} for the lattice QCD study of the ground state mesons. For the radial excited hadrons, see Ref.  \cite{Green2004,Burch2006,Burch2006a,Dudek2009,Edwards2011,Liu2012}.

My purpose herein is to study the radial excited states via a continuum QCD approach, the Dyson-Schwinger equation and Bethe-Salpeter equation (DSBSE) approach. Many interesting properties of the excited states have been obtained in this approach, for example, the radial excited pseudoscalar mesons decouple from the axial-vector current in the chiral limit \cite{Holl2004}. The electromagnetic properties and the chiral symmetry restoration of the radial excited states were also studied \cite{Holl2005,Wagenbrunn2007}. It is showed qualitatively that the radial excited mesons are more sensitive to the interaction at large distance though a realistic prediction for the excited mesons from the rainbow ladder (RL) approximation is failed \cite{Krassnigg2008,Qin2012}.

In this paper, I focus on the the first radial excited heavy pseudoscalar and vector mesons, i.e. $\eta_c(2S)$, $\psi(2S)$, $B_c(2S)$, $B^*_c(2S)$, $\eta_b(2S)$ and $\varUpsilon(2S)$. Of these mesons, $\psi(2S)$ and $\varUpsilon(2S)$ were discovered more than 20 years ago \cite{Abe1995,Abe1997}, and were extensively studied in the following years. Even so, the decay width of $\eta_b(2S)$ is not well determined. Then $\eta_c(2S)$ was discovered in 2004 \cite{Aubert2004} and $\eta_b(2S)$ in 2012 \cite{Mizuk2012}. The $B_c$ mesons are much more difficult to observe due to the small production cross sections. Until recently $B_c(2S)$ and $B^*_c(2S)$ were measured by CMS and LHCb collaboration \cite{Sirunyan2019,Aaij2019}, yet the mass of $B^*_c(2S)$ is not determined  precisely. Therefore a systematic theoretical study of these mesons is of significant meaning.

I and my collaborators studied these excited mesons in the RL approximation in Ref. \cite{Chang2020}. Therein the one gluon exchange effective interaction between the quark and antiquark is fixed by the ground state mesons \cite{Chen2019}. The effective interaction in Ref. \cite{Chen2019} is universal for all the light, heavy-light and heavy ground pseudoscalar and vector mesons. It could be extended to the ground state scalar and axial vector mesons \cite{Chen2021}. When applying the same interaction to the radial excited heavy mesons, we found that the spectrum is 1\% lower than the available experiment value. The leptonic decay constant is lower by 12\% for $\varUpsilon(2S)$ and 42\% for $\psi(2S)$. These tell us that the radial excited mesons do not share the same effective interaction with the ground state mesons.

What the effective interaction should be? The proper interaction should not only produce the right spectrum, but also the right wave function of the mesons. The leptonic decay constant (see Eq. (\ref{eq:f0-}) and Eq. (\ref{eq:f1-})) describes the quark antiquark annihilation inside the meson and thus is the simplest quantity related to the meson wave function. So herein I refix the effective interaction by the masses and leptonic decay constants of $\varUpsilon(2S)$ and $\psi(2S)$. Then masses and leptonic decay constants of the other four resonances are calculated. It turns out that all the first radial excited heavy pseudoscalar and vector mesons could be described in a universal effective interaction. In section \ref{sec:model}, I introduce the framework of the DSBSE approach in the RL approximation. The details of the calculation and the results are expounded in section \ref{sec:results}. At last, summary and conclusions
are given in section \ref{sec:conclusion}.

\section{The model}\label{sec:model}
\noindent

The framework has been introduced in the previous works \cite{Chen2019,Chang2020,Chen2021}. It is recapitulated here for convenience. In the RL approximation, the quark propagator is solved by the Gap equation \cite{Dyson1949,Schwinger1951}:
\begin{equation}\label{eq:DSERL}
 S_f^{-1}(k) = Z_2 (i\gamma\cdot k + Z_m m_f) 
+ \frac{4}{3} (Z_2)^2 \int^\Lambda_{d q} \tilde{D}^f_{\mu\nu}(l)\gamma_\mu S_{f}(q)\gamma_\nu, 
\end{equation}
where $f=\{u,d,s,c,b,t\}$ represents the quark flavor. $S_{f}(k)$ is the quark propagator, which can be decomposed as $S_{f}(k) = \frac{Z_f(k^2)}{i\slashed{k} + M_f(k^2)}$. $Z_f(k^2)$ is the quark dressing function and $M_f(k^2)$ the quark mass function. $l=k-q$. $m_f$ is the current quark mass. $\tilde{D}^f_{\mu\nu}$ represents the effective interaction in the self energy of $f$-quark. $Z_2$ and $Z_m$ are the renormalisation constants of the quark field and the quark mass respectively.
$\int^\Lambda_{d q}=\int ^{\Lambda} d^{4} q/(2\pi)^{4}$ stands for a Poincar$\acute{\text{e}}$ invariant regularized integration, with $\Lambda$ the regularization scale.

A meson is qualified by the Bethe-Salpeter amplitude (BSA), $\Gamma^{fg}(k;P)$, with $k$ and $P$ the relative and the total momentum of the meson. $P^2 = -M^2_{fg}$ and $M_{fg}$ is the mass of the meson with quark flavor $(f,g)$. The BSA is solved by the Bethe-Salpeter equation (BSE) \cite{Salpeter1951,Roberts2017}
\begin{equation}\label{eq:BSERL}
  \big{[} \Gamma^{fg}(k;P)  \big{]}^{\alpha}_{\beta}  =  - \int^\Lambda_{d q} \frac{4}{3}(Z_{2})^{2} \tilde{D}^{fg}_{\mu\nu}(l) [\gamma_{\mu}^{}]^{\alpha}_\sigma [\gamma_{\nu}]^\delta_\beta \big{[} \chi^{fg}(q;P)  \big{]}^{\sigma}_{\delta} ,
\end{equation}
where $\alpha$, $\beta$, $\sigma$ and $\delta$ are the Dirac indexes.
$\chi^{fg}(q;P) = S_{f}(q_{+}) \Gamma^{fg}(q;P) S_{g}(q_{-})$ is the wave function, $q_{+} = q + \eta P/2$, $q_{-} = q - (1-\eta) P/2$, $\eta$ is the partitioning parameter describing the momentum partition between the quark and antiquark and dosen't affect the physical observables. $\tilde{D}^{fg}_{\mu\nu}$ represents the effective interaction between the quark and the antiquark. The leptonic decay constant of the pseudoscalar meson, $f_{0^{-}}$, is obtained by
\begin{equation}\label{eq:f0-}
f_{0^{-}}P_{\mu} = Z_{2} N_{c} \;\text{tr} \! \int^{\Lambda}_{d q} \! \gamma_{5}^{} \gamma_{\mu}^{} S_f(q_+)\Gamma^{fg}_{0^{-}}(q;P)S_g(q_-).
\end{equation}
$N_c = 3$, is the color number. ``$\text{tr}$" represents the trace of the Dirac index. The leptonic decay constant of the vector meson, $f_{1^{-}}$, is analogue
\begin{equation}\label{eq:f1-}
 f_{1^{-}}M_{1^-} = Z_{2} N_{c} \;\text{tr} \! \int^{\Lambda}_{d q} \! \gamma_{\mu}^{} S_f(q_+)\Gamma^{fg,\mu}_{1^{-}}(q;P)S_g(q_-).
\end{equation}

$\tilde{D}^{f}_{\mu\nu}$ and $\tilde{D}^{fg}_{\mu\nu}$ are decomposed as $\tilde{D}^{f}_{\mu\nu}(l) = \left(\delta_{\mu\nu}-\frac{l_{\mu}l_{\nu}}{l^{2}}\right)\mathcal{G}^f(l^2)$ and $\tilde{D}^{fg}_{\mu\nu}(l) = \left(\delta_{\mu\nu}-\frac{l_{\mu}l_{\nu}}{l^{2}}\right)\mathcal{G}^{fg}(l^2)$. The dressing functions $\mathcal{G}^f(l^2)$ and $\mathcal{G}^{fg}(l^2)$ are modeled as
\begin{eqnarray}\label{eq:gluonmodel}
  \mathcal{G}^f(s) 	&=& \mathcal{G}^f_{IR}(s) + \mathcal{G}_{UV}(s),\\\label{eq:gluonInfrared}
  \mathcal{G}^f_{IR}(s) &=& 8\pi^2\frac{D_f^2}{\omega_f^4} e^{-s/\omega_f^2},\\\label{eq:gluonfmodel}
  \mathcal{G}^{fg}(s) 	&=& \mathcal{G}^{fg}_{IR}(s) + \mathcal{G}_{UV}(s),\\\label{eq:gluonfInfrared}
  \mathcal{G}^{fg}_{IR}(s) 	&=& 8\pi^2\frac{D_f}{\omega_f^2}\frac{D_g}{\omega_g^2} e^{-s/(\omega_f\omega_g)},\\\label{eq:gluonUltraviolet}
  \mathcal{G}_{UV}(s) 	&=& \frac{8\pi^{2} \gamma_{m}^{} \mathcal{F}(s)}{\text{ln}[\tau+(1+s/\Lambda^{2}_{QCD})^2]},
\end{eqnarray}
where $s=l^2$. $\mathcal{G}^f_{IR}(s)$ and $\mathcal{G}^{fg}_{IR}(s)$ are the infrared interaction responsible for hadron properties. $\omega_f$ represents the interaction width in the momentum space, and $D_f$ is the infrared strength. $\mathcal{G}_{UV}(s)$ keeps the one-loop perturbative QCD limit in the ultraviolet.
$\mathcal{F}(s)=[1 - \exp(-s^2/[4m_{t}^{4}])]/s$, $\gamma_{m}^{}=12/(33-2N_{f})$, with $m_{t}=1.0 \textmd{ GeV}\,$, $\tau=e^{10} - 1$, $N_f=5$, and $\Lambda_{\text{QCD}}=0.21 \textmd{ GeV}\,$. This model turned out to be successful for all the ground state pseudoscalar and vector mesons, from heavy, heavy-light to light mass scales. It has been extended to the scalar and axial vector mesons \cite{Chen2021}. These successfulness support that Eq. (\ref{eq:gluonfInfrared}) contains the proper flavor dependence. Though the same interaction should not apply to the radial excited mesons if high precision is required, as stated in section \ref{sec:intro}. 

To mimic the interesting difference between the radial excited states and the ground states, Eq. (\ref{eq:gluonfInfrared}) is changed into
\begin{equation}\label{eq:gluonfInfrarednew}
   \mathcal{G}^{fg}_{IR}(s) = 8\pi^2\frac{\eta_f D_f}{\omega_f^2}\frac{ \eta_ g D_g}{\omega_g^2} e^{-s/(\alpha_f\omega_f \alpha_g\omega_g)}.
\end{equation}
While $\omega_f$ and $D_f$ ($f=\{c,b\}$) are kept unchanged from Ref. \cite{Chen2019}, $\alpha_f$ and $\eta_f$ ($f=\{c,b\}$) are free parameters. $\alpha_f$ presents the changing of the interaction width and $\eta_f$ the infrared strength. The four free parameters, $\alpha_c$, $\alpha_b$, $\eta_c$ and $\eta_b$, are fixed by fitting the masses and leptonic decay constants of $\psi(2S)$ and $\varUpsilon(2S)$ ($M_{\psi(2S)}$, $M_{\varUpsilon(2S)}$, $f_{\psi(2S)}$, $f_{\varUpsilon(2S)}$) to the experiment values. Where the expriment value of $f_{\psi(2S)}$ and $f_{\varUpsilon(2S)}$ are extracted from the branch decay width of the vector meson to $e^+ e^-$. The values are $f_{\psi(2S)} = -0.208\textmd{ GeV}$, $f_{\varUpsilon(2S)} = -0.352\textmd{ GeV}$  \cite{Chang2020}. Then masses and leptonic decay constants of $\eta_c(2S)$, $B_c(2S)$, $B^*_c(2S)$, $\eta_b(2S)$ are calculated with the refixed interaction.

\section{Outputs of the model}\label{sec:results}
\noindent

Before discussing the results, we should first explain the solving process of the BSE, Eq. (\ref{eq:BSERL}). Eq. (\ref{eq:BSERL}) is solved as a $P^{2}$-dependent eigenvalue problem,

\end{multicols}

\begin{center} 
 \includegraphics[width=0.45\textwidth]{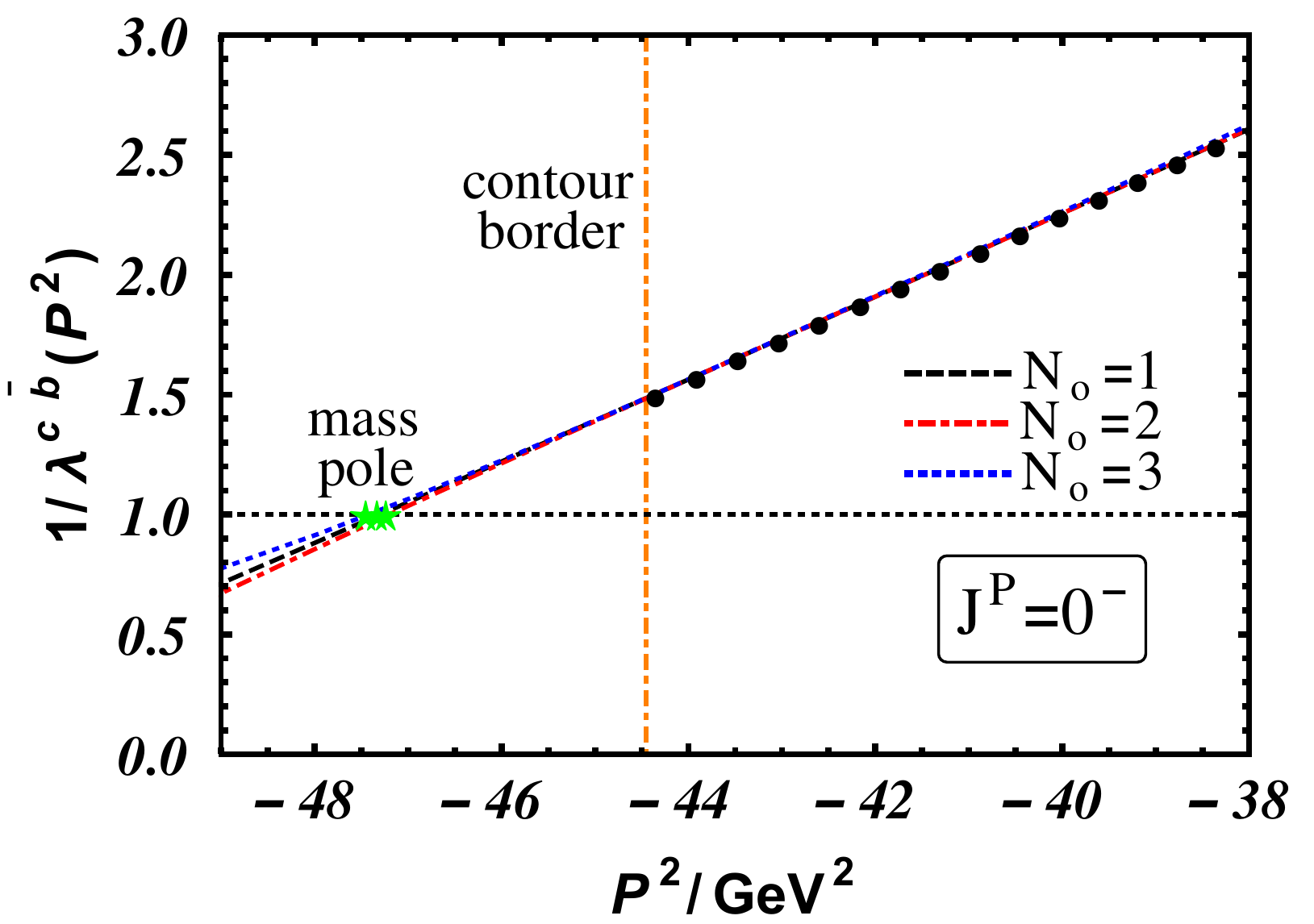}
 \includegraphics[width=0.47\textwidth]{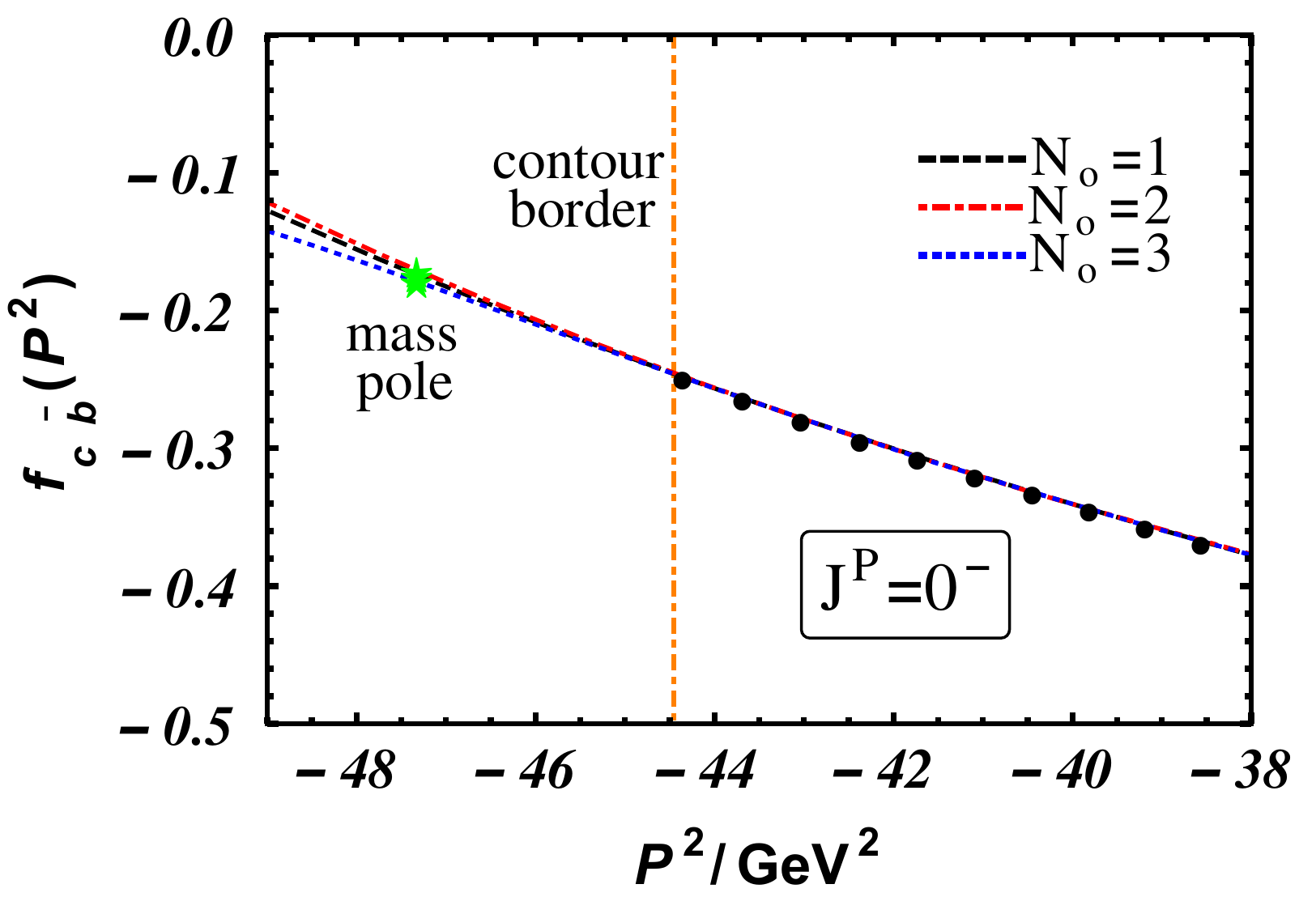}
 \includegraphics[width=0.45\textwidth]{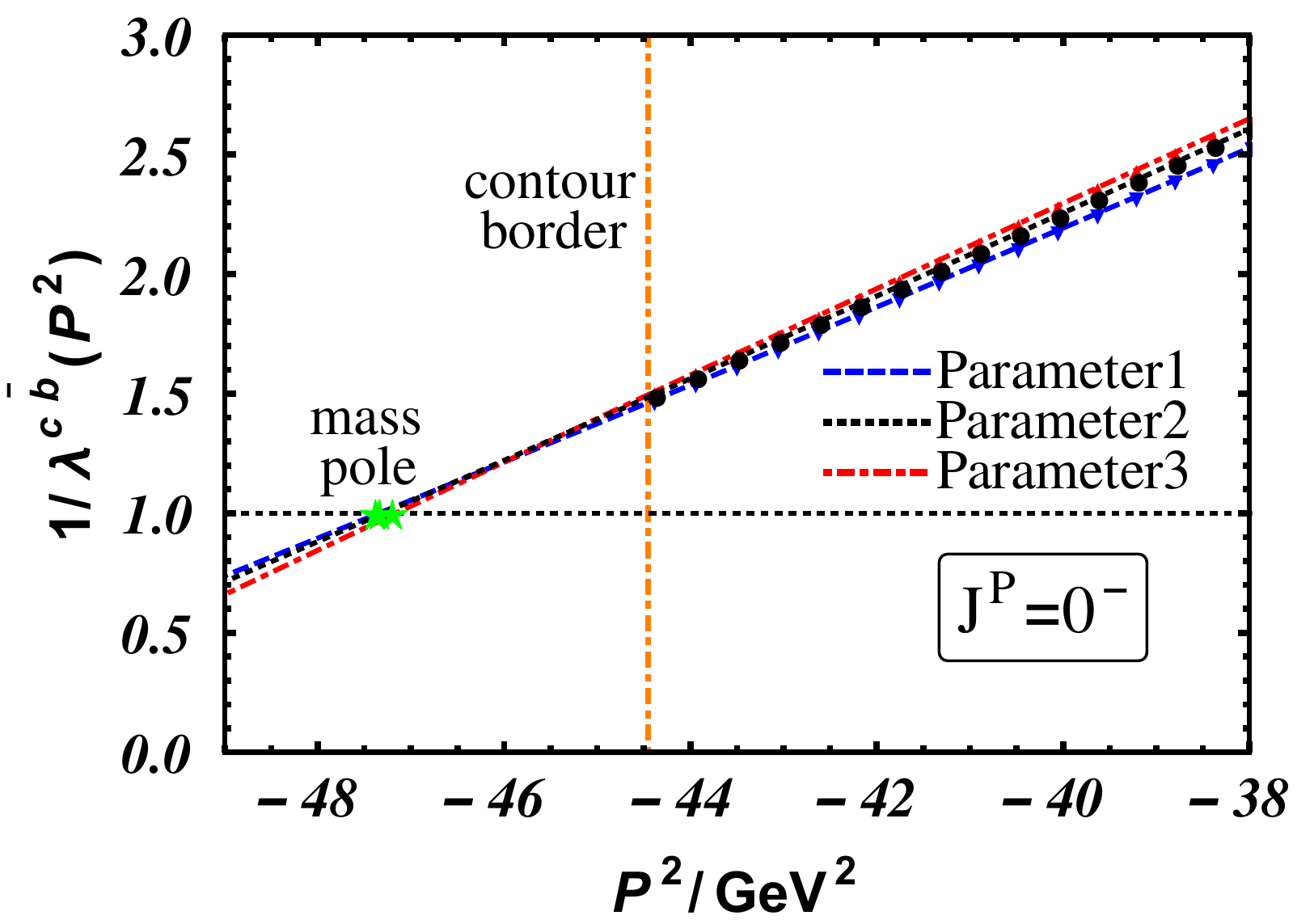}
 \includegraphics[width=0.47\textwidth]{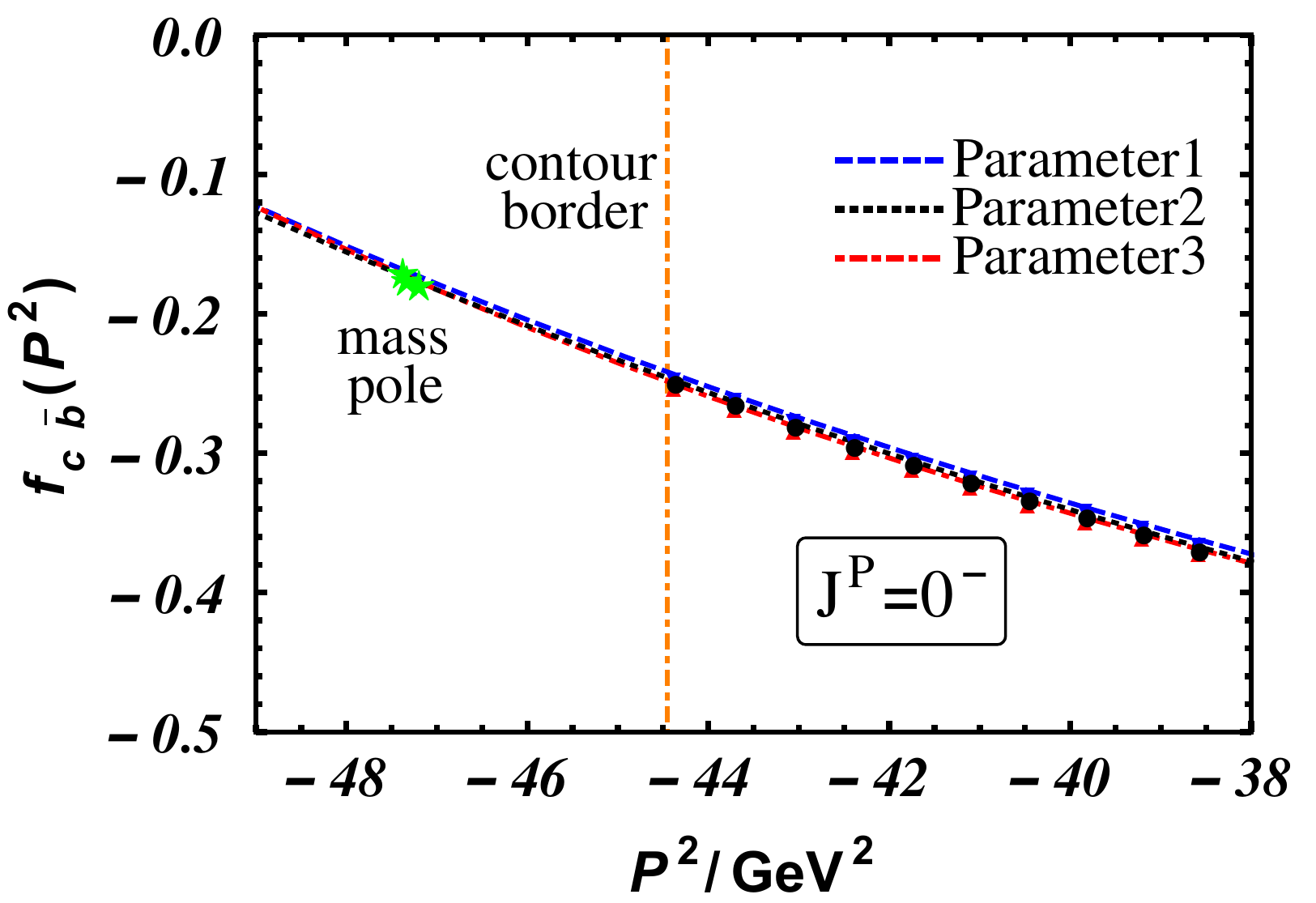}
\figcaption{\label{fig:massf-ext} The extrapolation of the eigenvalue, $\lambda(P^2)$, and the leptonic decay constant, $f(P^2)$, of $B_c(2S)$. The upper two figures show the extrapolation with $N_o = 1, 2, 3$ in Eq. (\ref{eq:fitlambda}) and Eq. (\ref{eq:fitf}). The lower two figures show the results using three different groups of parameters that are listed in Tab. \ref{tab:parameters} in the appendix. The vertical dot-dashed orange line is the contour border on the right of which the direct calculation can be applied. The green stars present the location of the meson masses.}
\end{center}

\begin{equation}\label{eq:BSERL2}
 \lambda^{fg}(P^2) \big{[} \Gamma^{fg}(k;P)  \big{]}^{\alpha}_{\beta}  =  - \int^\Lambda_{d q} \frac{4}{3}(Z_{2})^{2} \tilde{D}^{fg}_{\mu\nu}(l) [\gamma_{\mu}^{}]^{\alpha}_\sigma [\gamma_{\nu}]^\delta_\beta \big{[} \chi^{fg}(q;P)  \big{]}^{\sigma}_{\delta}.
\end{equation}

\begin{multicols}{2}

The meson mass is determined by $\lambda^{fg}(P^2=-M^2_{fg})=1$. However, due to the singularity of the quark propagators in the complex momentum plane, there is an lower bound value of $P^2$ in Eq. (\ref{eq:BSERL2}). Only when $P^2 > -M_{max}^2$, Eq. (\ref{eq:BSERL2}) is solvable. $M_{max}^2$ defines the contour border of the calculable region of the quark propagators. See the appendix of Ref. \cite{Hilger2015} for this problem.

As the masses of the radial excited mesons are beyond the contour border, i.e. $M_{fg} > M_{max}$, the following form is used to fit $\lambda^{fg}(P^2)$,
\begin{equation}\label{eq:fitlambda}
 \frac{1}{\lambda^{fg}(P^2)} = \frac{1 + \sum^{N_{o}}_{n=1}\, a_{n} (P^{2} + s_{0}^{})^n}{1 + \sum^{N_{o}}_{n=1}\, b_{n} (P^{2} + s_{0}^{})^n} \, ,
\end{equation}
where $N_{o}$ is the order of the series, $s_{0}^{}$, $a_{n}$ and $b_{n}$ are the parameters to be determined by the least square method. And $f_{fg}(P^2)$ is fitted by
\begin{equation}\label{eq:fitf}
 f_{fg}(P^2) = \frac{f_0 + \sum^{N_o}_{n=1}\, c_n (P^2+M^2)^n}{1 + \sum^{N_o}_{n=1}\, d_n (P^2+M^2)^n},
\end{equation}
where $f_{0}^{}$, $c_{n}$ and $d_{n}$ are parameters, $M^{2}$ is the square of the mass.
The physical value of the decay constant is $f_{fg}(-M^2) = f_{0}^{}$. This extrapolation scheme has been used in Ref. \cite{Chang2020,Chen2021}. The error is controllable as long as the physical mass of the meson is not too larger than $M_{max}$. The extrapolation results are illustrated by $B_c(2S)$, the two upper figures of Figure \ref{fig:massf-ext}. $N_o = 1, 2, 3$ are considered in practice, and the differences between the three cases are estimated as the error from extrapolation.

\end{multicols}

\newpage

\begin{center}
\tabcaption{\label{tab:massf} Masses and leptonic decay constants of the first radial excited heavy pseudoscalar and vector mesons (in GeV). The normalization convention $f_\pi = 0.093\textmd{ GeV}$ is used for the leptonic decay constants. $M^{\textmd{DSE}}_{q\bar{q}'}$ and $f^{\textmd{DSE}}_{q\bar{q}'}$ are the Dyson-Schwinger equation results, of which the first error is from the extrapolation, and the second is from varying the parameters. The underlined values are used to fit the parameters $\alpha_{f,g}$ and $\eta_{f,g}$ in Eq. (\ref{eq:gluonfInfrarednew}). $M^{\textmd{expt.}}_{q\bar{q}'}$ are the experiment values \cite{Zyla2020}. See the text for the experiment values of $f_{\psi(2S)}$ and $f_{\varUpsilon(2S)}$. The results in Ref \cite{Lakhina2006,Kiselev2001,Aliev2019,Soni2018} are devided by $\sqrt{2}$ to compare with mine. Ref. \cite{Lakhina2006} are the nonrelativistic quark model results, cited in the case of ``BGS log" with $\Lambda = 0.25\textmd{ GeV}$ and $\Lambda = 0.40\textmd{ GeV}$ therein. Ref. \cite{Kiselev2001} is the static potential model results. Ref. \cite{Aliev2019} is a QCD sum rule estimation and Ref. \cite{Soni2018} is a nonrelativistic Cornell potential model result.
}
\centering

\renewcommand\arraystretch{1.2}
\begin{tabular}{p{3em}<{\centering}|p{1.6em}<{\centering}|p{5em}<{\centering}|p{3em}<{\centering}|p{5.5em}<{\centering}|p{3em}<{\centering}|p{8.2em}<{\centering}|p{3em}<{\centering}}
\hline    
 meson & {$J^{\textmd{PC}}$}    &	{$M^{\textmd{DSE}}_{c\bar{c}}$}	& {$M^{\textmd{expt.}}_{c\bar{c}}$} &	{$f^{\textmd{DSE}}_{c\bar{c}}$}	& {$f^{\textmd{expt.}}_{c\bar{c}}$}& $|f|$\textsuperscript{\cite{Lakhina2006}}& $|f|$\textsuperscript{\cite{Soni2018}}\\
\hline
$\eta_c(2S)$ & $0^{-+}$ &  3.618(25)(3) & 3.638 &  -0.158(8)(4) & -- & 0.170 $\sim$ 0.172 & 0.197\\
$\psi(2S)$ & $1^{--}$ & \underline{3.686}(21)(0) & 3.686 & \underline{-0.208}(5)(0) & -0.208& 0.207$\sim$ 0.216 &0.182\\
\hline
\end{tabular}

\begin{tabular}{p{3em}<{\centering}|p{1.6em}<{\centering}|p{5em}<{\centering}|p{3em}<{\centering}|p{5.5em}<{\centering}|p{3em}<{\centering}|p{3.5em}<{\centering}|p{3.5em}<{\centering}|p{3em}<{\centering}}
\hline    
 meson & {$J^{\textmd{P}}$}    &	{$M^{\textmd{DSE}}_{c\bar{b}}$}	& {$M^{\textmd{expt.}}_{c\bar{b}}$} &	{$f^{\textmd{DSE}}_{c\bar{b}}$}	& {$f^{\textmd{expt.}}_{c\bar{b}}$} & $|f|$ \hspace*{-0.3em}\textsuperscript{\cite{Kiselev2001}} & $|f|$ \hspace*{-0.3em}\textsuperscript{\cite{Aliev2019}}& $|f|$\textsuperscript{\cite{Soni2018}}\\
\hline
$B_c(2S)$ & $0^{-}$ & 6.874(9)(6) & 6.872 & -0.174(5)(4) & -- &$0.198(35)$ & 0.304(14) & 0.251\\
$B^*_c(2S)$ & $1^{-}$ & 6.926(12)(6) & --    & -0.216(9)(4) & -- & -- & 0.325(14) & 0.252\\
\hline
\end{tabular}

\begin{tabular}{p{3em}<{\centering}|p{1.6em}<{\centering}|p{5em}<{\centering}|p{3em}<{\centering}|p{5.5em}<{\centering}|p{3em}<{\centering}|p{8.2em}<{\centering}|p{3em}<{\centering}}
\hline    
 meson & {$J^{\textmd{PC}}$}    &{$M^{\textmd{DSE}}_{b\bar{b}}$}	& {$M^{\textmd{expt.}}_{b\bar{b}}$} &	{$f^{\textmd{DSE}}_{b\bar{b}}$}	& {$f^{\textmd{expt.}}_{b\bar{b}}$}& $|f|$\textsuperscript{\cite{Lakhina2006}}& $|f|$\textsuperscript{\cite{Soni2018}}\\
\hline
$\eta_b(2S)$& $0^{-+}$ & 9.989(13)(3) & 9.999 & -0.345(6)(1) & -- &0.291$\sim$ 0.299 & 0.367\\
$\varUpsilon(2S)$ & $1^{--}$ & \underline{10.023}(11)(0) & 10.023 & \underline{-0.352}(4)(0) & -0.352 & 0.336$\sim$ 0.350 & 0.367\\
\hline
\end{tabular}
\end{center}

\begin{multicols}{2}

Three groups of $\omega_f$ and $D_f$ are used, which are the same as Ref. \cite{Chen2019}. Different groups of parameters correspond to different interaction width. In each case, $\alpha_f$ and $\eta_f$ ($f=\{c,b\}$), are fixed by fitting the masses and leptonic decay constants of $\psi(2S)$ and $\varUpsilon(2S)$ to the experiment values. Then masses and leptonic decay constants of $\eta_c(2S)$, $B_c(2S)$, $B^*_c(2S)$, $\eta_b(2S)$ are outputted. The differences of the outputs are considered as the errors from varing the parameters. This is illustrated by $B_c(2S)$, the two lower figures of Figure \ref{fig:massf-ext}.

The masses and leptonic decay constants of $\eta_c(2S)$, $\psi(2S)$, $B_c(2S)$, $B^*_c(2S)$, $\eta_b(2S)$, $\varUpsilon(2S)$ herein (the DSE results) and the available experiment values (the expt. results) are listed in Table \ref{tab:massf}. For the DSE results, the first error is from the extrapolation, and the second is from varying the parameters. The largest mass deviation of the DSE results from the experiment values is $20\textmd{ MeV}$, in the case of $\eta_c(2S)$. The mass of $B^*_c(2S)$ is not well determined experimentally, due to the absence of the $B^*_c(1S)$ mass. Herein I provide the most precise prediction of the $B^*_c(2S)$ mass. The systematic error of the spectrum is estimated to be $20\textmd{ MeV}$.

There are no experiment values for $f_{\eta_c(2S)}$, $f_{B_c(2S)}$, $f_{B^*_c(2S)}$, $f_{\eta_b(2S)}$ yet. I list the leptonic decay constants from other models in Table \ref{tab:massf}. In this article, the normalization convention $f_\pi = 0.093\textmd{ GeV} \approx 0.131/\sqrt{2} \textmd{ GeV}$ is used for the leptonic decay constants, so the results in Ref \cite{Lakhina2006,Kiselev2001,Aliev2019} are devided by $\sqrt{2}$ to compare with mine. Ref. \cite{Lakhina2006} are the nonrelativistic quark model results, cited in the case of ``BGS log" with $\Lambda = 0.25\textmd{ GeV}$ and $\Lambda = 0.40\textmd{ GeV}$ therein. Ref. \cite{Kiselev2001} is the static potential model results and Ref. \cite{Aliev2019} is a QCD sum rule estimation. There are some other model estimation of the leptonic decay constant of $\eta_c(2S)$, $\psi(2S)$, $\eta_b(2S)$, $\varUpsilon(2S)$ recently \cite{Negash2016,Azhothkaran2020}.
However, Ref. \cite{Lakhina2006} is the most consistent one comparing with mine. There are much fewer studies of $B_c(2S)$ and $B^*_c(2S)$ leptonic deacy constants. Results from Ref. \cite{Kiselev2001} is consistent with mine. But results from Ref. \cite{Aliev2019} seem impossible from the viewpoint herein. Ref. \cite{Soni2018} calculated all the S-wave heavy meson spectrum and the leptonic decay constants via nonrelativistic Cornell potential model. The bottomonium decay constants therein are consistent with mine, while there are deviations for charmonium and the $B_c$ mesons.
 
Anyway the reasonableness of my results could be justified by the following three facts:
\begin{enumerate}
 \item the RL approximation is suffcient for the pseudoscalar and vector mesons;
 \item the interaction pattern Eq. (\ref{eq:gluonfmodel}), Eq.(\ref{eq:gluonfInfrarednew}) and Eq. (\ref{eq:gluonUltraviolet}) contains the proper flavor dependence, so the $B_c$ mesons share this universal interaction form with the charmonium and the bottomonium;
 \item the interaction is refixed by the experimental value of masses and leptonic decay constants of $\varUpsilon(2S)$ and $\psi(2S)$, so it's the realistic interaction for the radial excited mesons.
\end{enumerate}

At last, let's discusse the effective interaction between the quark and antiquark in the radial excited mesons. It is characterized by the dressing function $\mathcal{G}^{fg}$. The dressing functions, $\mathcal{G}^{c\bar{c}}(k^2)$, $\mathcal{G}^{c\bar{b}}(k^2)$ and $\mathcal{G}^{b\bar{b}}(k^2)$, for the ground states and the first radial excited states are depicted in Fig. \ref{fig:dressingG}. We can see that the dressing functions of the radial excited mesons are smaller in the region $k^2 \lesssim 0.7 \sim 1.1\textmd{ GeV}^2$ and larger in the region $k^2 \gtrsim 0.7 \sim 1.1\textmd{ GeV}^2$. This can be understood as: the energy of quarks in the radial excited mesons are larger, so the soft interaction declines and the hard interaction raises. This feature also applys to the light mesons. As the light excited meson mass is farther from the contour border, the nature extrapolation Eq. (\ref{eq:fitlambda}) and Eq. (\ref{eq:fitf}) has a larger error. This problem is reserved for future studies.
 
\begin{center}
 \includegraphics[width=0.46\textwidth]{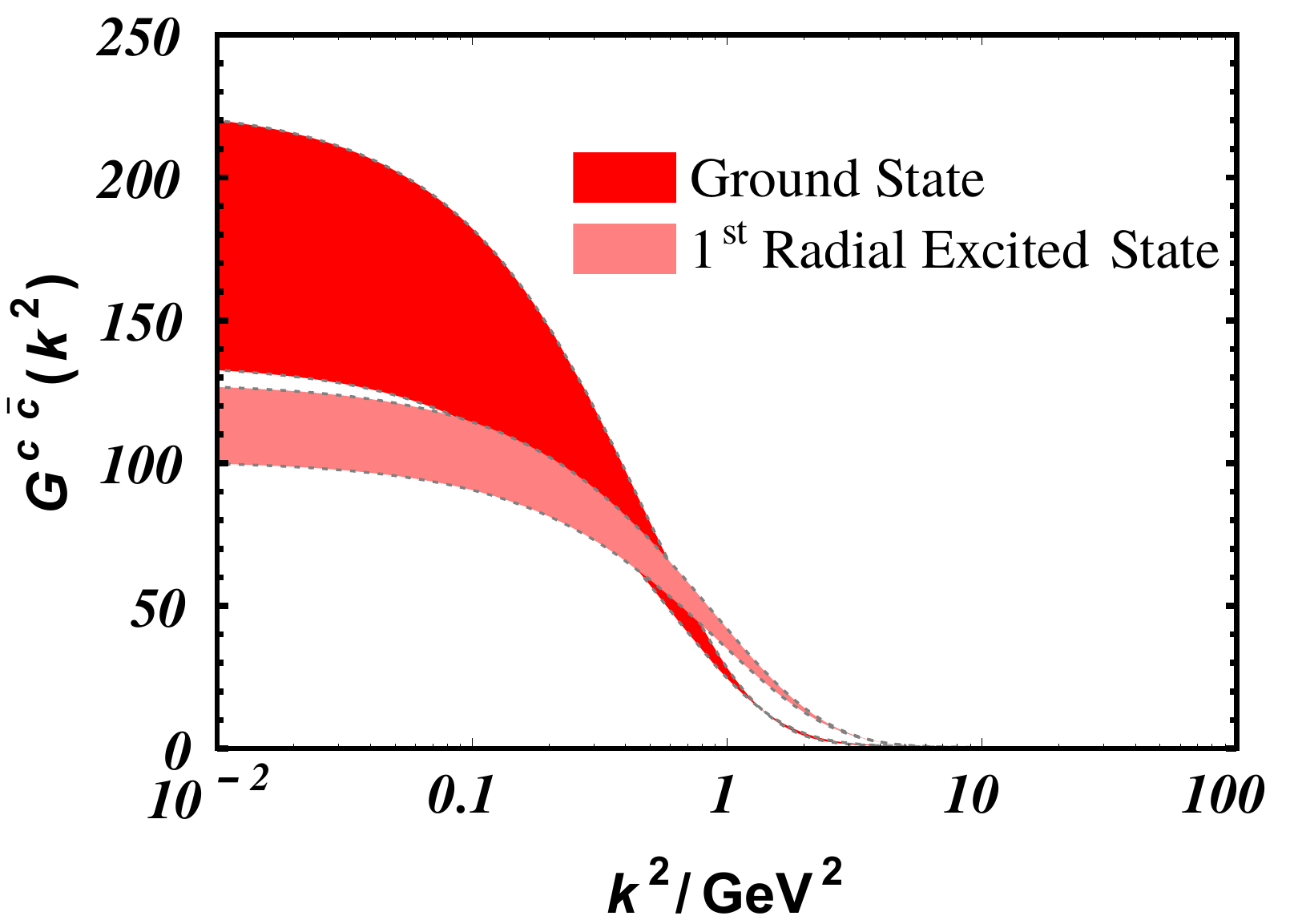}
 \includegraphics[width=0.46\textwidth]{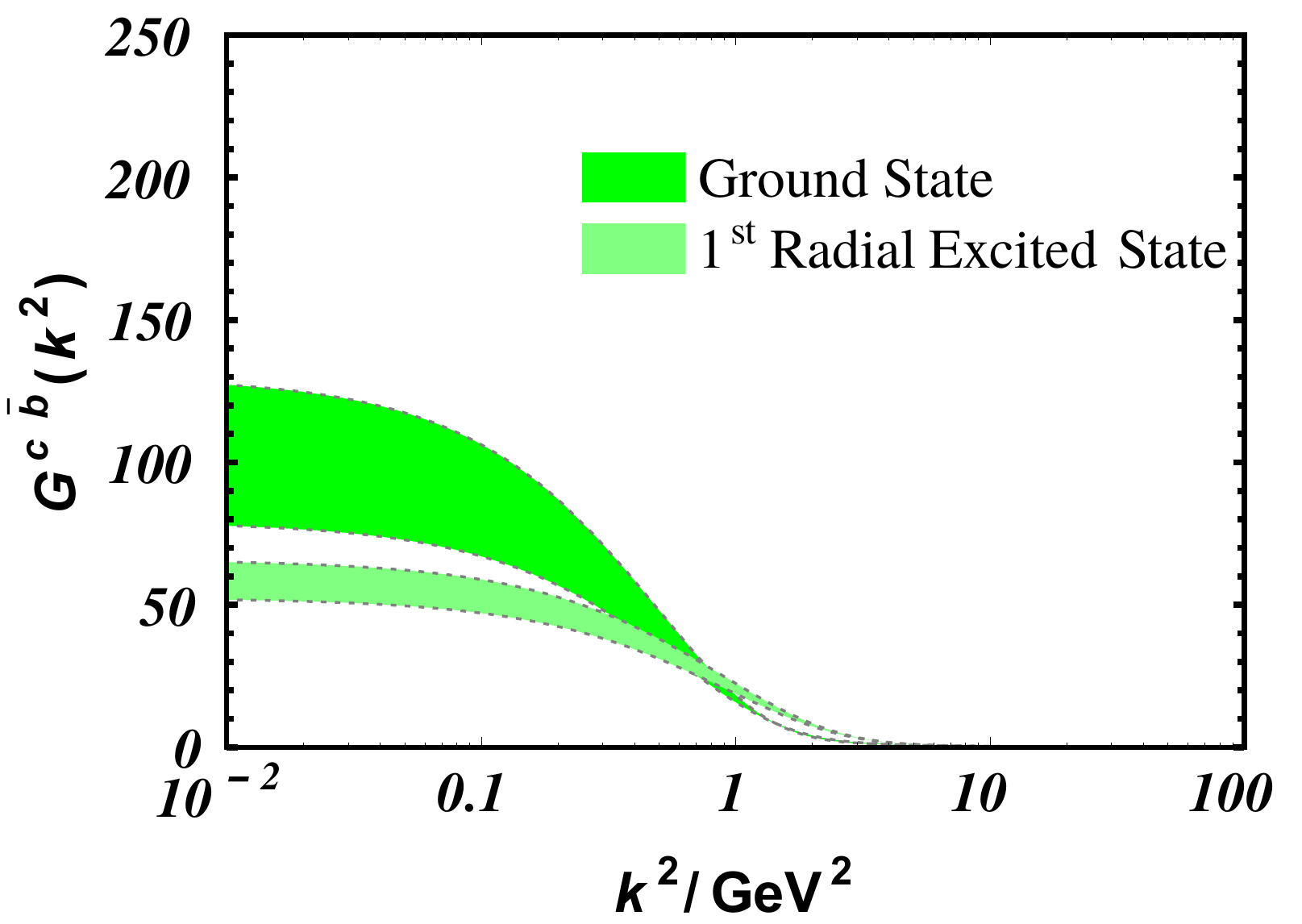}
 \includegraphics[width=0.46\textwidth]{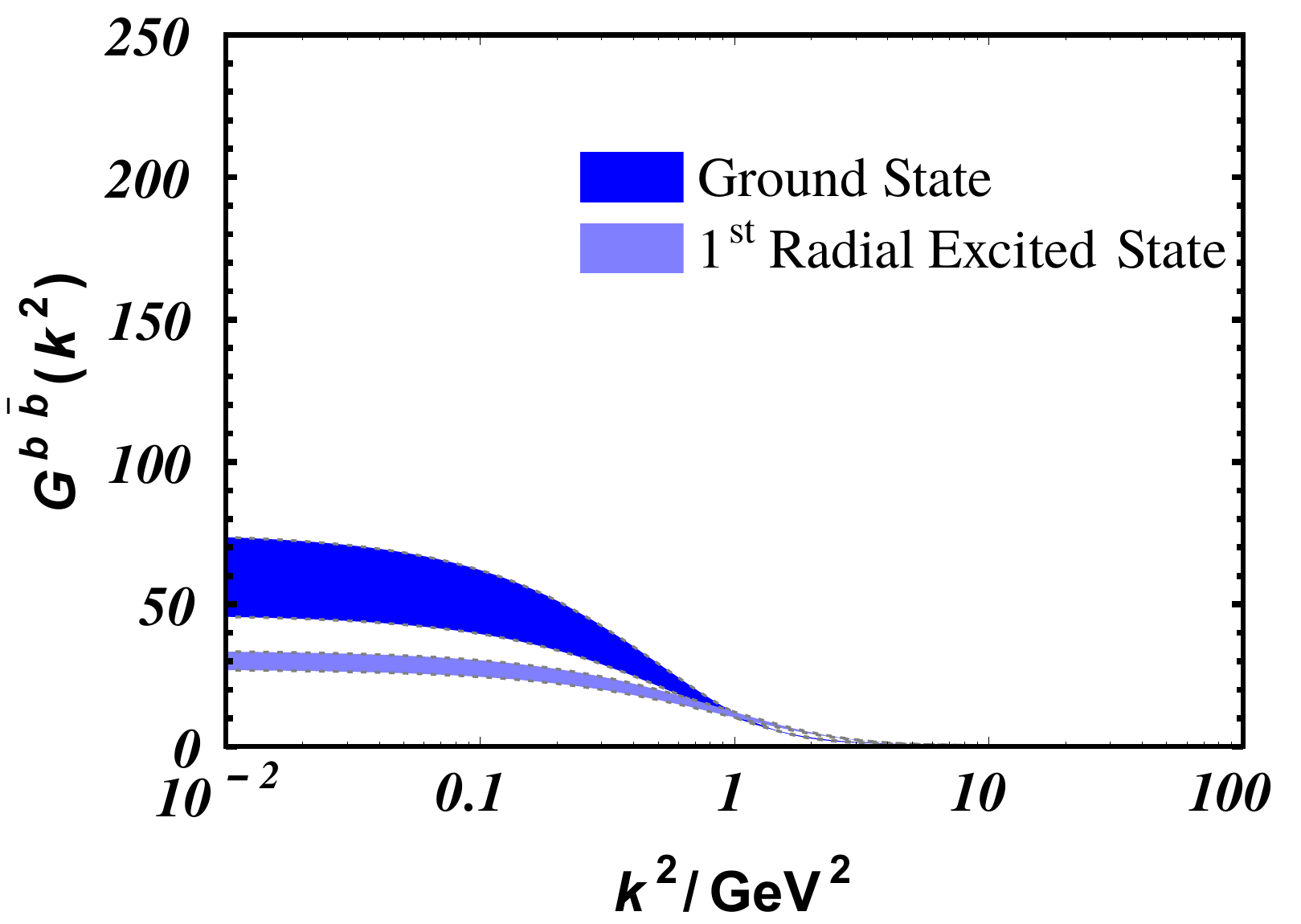}
\figcaption{\label{fig:dressingG} The effective interaction dressing functions of the ground states (Eq. (\ref{eq:gluonfmodel}), Eq.(\ref{eq:gluonfInfrared}) and Eq. (\ref{eq:gluonUltraviolet})) and the first radial excited states (Eq. (\ref{eq:gluonfmodel}), Eq.(\ref{eq:gluonfInfrarednew}) and Eq. (\ref{eq:gluonUltraviolet})). The region boundary is defined by Parameter-1 and Parameter-3 in Tab. \ref{tab:parameters} in the appendix.}
\end{center}

\section{Summary and conclusion}\label{sec:conclusion}
\vspace{0.1cm}

In summary, the first radial excited heavy pseudoscalar and vector mesons ($\eta_c(2S)$, $\psi(2S)$, $B_c(2S)$, $B^*_c(2S)$, $\eta_b(2S)$, $\varUpsilon(2S)$) are studied in the Dyson-Schwinger equation and Bethe-Salpeter equation approach. The interaction is refixed by fitting the masses and leptonic decay constants of $\psi(2S)$ and $\varUpsilon(2S)$ to the expriment values. With the interaction well determined, a precise prediction of the masses and leptonic decay constants of $\eta_c(2S)$, $B_c(2S)$, $B^*_c(2S)$ and $\eta_b(2S)$ are given. My study also shows that, comparing with the ground states, the soft part of the effective interaction in the excited states declines and the hard part raises.



\section{Appendix A}\label{sec:appendixA}

The three groups of parameters correspond to $\omega_u = 0.45, 0.50, 0.55 \textmd{ GeV}$ are listed in Tab.\ref{tab:parameters}.
The quark mass $\bar{m}_f^{\zeta}$ is defined by
\begin{eqnarray}
 \bar{m}_f^{\zeta} &=& \hat{m}_f/\left[\frac{1}{2}\textmd{Ln}\frac{\zeta^2}{\Lambda^2_{\textmd{QCD}}}\right]^{\gamma_m},\\
 \hat{m}_f	 &=& \lim_{p^2 \to \infty}\left[\frac{1}{2}\textmd{Ln}\frac{p^2}{\Lambda^2_{\textmd{QCD}}}\right]^{\gamma_m} M_f(p^2),
\end{eqnarray}
with $\zeta$ the renormalisation scale, $\hat{m}_f$ the renormalisation-group invariant current-quark mass and $M_f(p^2)$ the quark mass function.

\begin{center}
 \tabcaption{\label{tab:parameters} Three groups of parameters correspond to $\omega_{u/d} = 0.45, 0.50, 0.55 \textmd{ GeV}$. $\bar{m}_f^{\zeta=2\textmd{GeV}}$, $\omega_f$ and $D_f$ are measured in GeV. $\alpha$ and $\eta$ are of unit 1.}
\centering
\renewcommand\arraystretch{1.2}
\begin{tabular}{c|c|c|c|c|c}
\hline\hline
\multirow{2}{*}{flavor} & \multirow{2}{*}{$\bar{m}_f^{\zeta=2\textmd{GeV}}$} & \multicolumn{4}{c}{Parameter-1} \\
\cline{3-6}
 & & $w_f $ & $D_f^2$ & $\alpha_f$ & $\eta_f$ \\
 \hline
 c & 1.17 & 0.690 & 0.645 & 1.360 & 0.755 \\
\hline
b & 4.97 & 0.722 & 0.258 & 1.323 & 0.671 \\
\hline
\end{tabular}

\begin{tabular}{c|c|c|c|c|c}
\hline
\multirow{2}{*}{flavor} & \multirow{2}{*}{$\bar{m}_f^{\zeta=2\textmd{GeV}}$} & \multicolumn{4}{c}{Parameter-2} \\
\cline{3-6}
 & & $w_f $ & $D_f^2$ & $\alpha_f$ & $\eta_f$ \\
 \hline
 c & 1.17 & 0.730 & 0.599 & 1.304 & 0.817 \\
\hline
b & 4.97 & 0.766 & 0.241 & 1.265 & 0.730 \\
\hline
\end{tabular}

\begin{tabular}{c|c|c|c|c|c}
\hline
\multirow{2}{*}{flavor} & \multirow{2}{*}{$\bar{m}_f^{\zeta=2\textmd{GeV}}$} & \multicolumn{4}{c}{Parameter-3} \\
\cline{3-6}
 & & $w_f $ & $D_f^2$ & $\alpha_f$ & $\eta_f$ \\
 \hline
 c & 1.17 & 0.760 & 0.570 & 1.265 & 0.865 \\
\hline
b & 4.97 & 0.792 & 0.231 & 1.225 & 0.766 \\
\hline\hline
\end{tabular}

\end{center}

\end{multicols}

\begin{multicols}{2}

\vspace{2mm}
\centerline{\rule{80mm}{0.1pt}}
\vspace{2mm}

%
%
%
%
%

\bibliography{res}

\end{multicols}

\clearpage

\end{document}